\documentclass[aps,prl,twocolumn]{revtex4}
\usepackage{amsmath,amssymb}
\usepackage{graphics}
\usepackage{graphicx}
\usepackage{color}

\begin{document}

\title{Intermittency of velocity time increments in turbulence.}

\author{L. Chevillard, S. G. Roux, E. L\' ev\^eque, N. Mordant${}^{(1)}$, J.-F. Pinton and A. Arn\'eodo}
\affiliation{Laboratoire de Physique, CNRS \& \'Ecole Normale Sup\'erieure
de Lyon,\\ 46 all\'ee d'Italie, F-69007 Lyon, France}
 \altaffiliation[${}^{(1)}$  Presently at ]{Laboratoire de Physique Statistique, CNRS \& \'Ecole Normale SupŽrieure, 24 rue Lhomond, 75005 Paris, France}

\begin{abstract}
We analyze the statistics of turbulent velocity fluctuations in
the time domain. Three cases are computed numerically and
compared:  (i) the time traces of Lagrangian fluid particles in a
(3D) turbulent flow (referred to as the {\it dynamic} case); (ii)
the time evolution of tracers  advected by a frozen turbulent
field (the {\it static} case), and  (iii) the evolution in time of
the velocity recorded at a fixed location in an evolving Eulerian
velocity field, as it would be measured by a local probe (referred
to as the {\it virtual probe} case). We observe that the static
case and the virtual probe cases share many properties with
Eulerian velocity statistics. The dynamic (Lagrangian) case is
clearly different; it bears the signature of the global dynamics
of the flow.
\end{abstract}

\maketitle

One of the distinctive feature of turbulence is the development of
extremely high fluctuation level at small scales. The probability
density functions (PDFs) of velocity increments are stretched at
small scales, while they are almost Gaussian at large scale where
energy is fed into the flow~\cite{Frisch}.  This evolution is
traditionally referred to as ``intermittency'' in turbulence
studies. Numerous studies have been devoted to the study of
intermittency in the spatial domain, analyzing velocity
differences between two points separated by a variable distance
$\mathbf{r}$. For instance, it is now well established that when
the velocity increments are computed along the distance
$\mathbf{r}$, the {\it longitudinal } velocity increments are
skewed and the structure functions have universal relative scaling
exponents in the inertial range~\cite{Arneodo,SheLev,Cas}, related
to the graininess of the dissipation.  When the increments are
related to changes in the velocity component perpendicular to the
distance $\mathbf{r}$, the corresponding {\it transverse}
structure functions display a different scaling, related to the
spatial distribution of vorticity~\cite{Transverse}. In contrast,
there has been much fewer studies of intermittency in the time
domain, {\it i.e.},  related to changes {\it in time} of the
velocity field. Two cases are of particular importance:  The
Lagrangian one, which pertains to the fluctuations in time of the
velocity of marked fluid particles, and the Eulerian one, where
one considers the variations in time of the velocity measured at a
fixed location in the flow. We consider of course the case where
turbulence develops in the absence of a mean flow, otherwise
Taylor's hypothesis trivially reduces the second case to a spatial
measurement~\cite{Tayloc}. Eulerian time fluctuations are
different because one expects, after Tennekes original
suggestion~\cite{Tenn}, that {\it sweeping} (the random advection
by large scale motion) plays an important role. In practice,
Eulerian intermittency in time is relevant for stationary bodies
exposed to turbulent flow conditions (atmospheric ones, for
instance).  Lagrangian intermittency, on the other hand, has
strong implications for processes such as mixing~\cite{LagMix},
combustion~\cite{LagCombustion}, or cloud
formation~\cite{LagRain}. Lagrangian data has recently been made
available in numerical simulations~\cite{PopeYeung,Biferale} and
experimental measurements~\cite{Voth98,OttMann,MordantPinton1}.
One  rather unexpected feature is the observation of long-time
correlations in the Lagrangian dynamics~\cite{MordantPinton2}. It
has been incorporated in recent stochastic models of Lagrangian
acceleration~\cite{Sawford,Reynolds} --- see also~\cite{Arin} for
a recent review. However, the full modelization of Lagrangian
velocity fluctuations is still in progress. Lagrangian statistics
are often computed from the advection of particles in a frozen
Eulerian field (to minimize computing overhead). We have thus
decided to compare this pseudo-Lagrangian statistics to that of
pure Lagrangian and Eulerian time fluctuations.

We study the problem numerically. We first compute the velocity changes of
fluid particles that are advected by a frozen 3D Eulerian velocity
field; that is, we consider a single snapshot of a converged
turbulent flow, and use it to advect fluid particles in this
frozen Eulerian field --- we call this the static case. Then we
compute the velocity variations of true Lagrangian particles, a
situation in which the Eulerian flow is also evolved in time
according to the Navier-Stokes equations --- this case is called
the dynamic case. Finally, we record the time evolution of the
Eulerian velocity at fixed locations of the computation domain, as
it would be measured by virtual velocity probes. We then perform a
comparative study of the intermittency characteristics of the
three velocity time signals. We show that the statistics of time
velocity increments depends on the situation considered:  the
static case and the virtual probes display intermittency features
that are reminiscent of Eulerian velocity statistics; the former
has multifractal spectra identical to the ones measured for 3D
Eulerian turbulence~\cite{Kestener}, while the later coincides
with traditional longitudinal Eulerian velocity increments
statistics~\cite{Frisch}.  The dynamic case, which is a true
Lagrangian measurement, displays significantly more intermittent
features.

The Navier-Stokes equations are integrated in a $256^3$ cubic
domain of size $2\pi$ by a parallel distributed memory
pseudo-spectral solver, using a second-order (in time) leap-frog
scheme. The large-scale kinetic-energy forcing is adjusted at each
time step by scaling the amplitudes of modes $1.5 \le k < 2.5$
uniformly (phases are left to fluctuate freely), so as to
compensate exactly the losses due to eddy dissipation in the
kinetic-energy budget~\cite{levkoud}. The Reynolds number based on
the Taylor microscale is $R_\lambda = 140$. The velocity root mean
square is 0.1214~m/s, the mean dissipation rate is $\epsilon =
0.0011 {\rm m}^2/{\rm s}^3$ and the kinematic viscosity is $\nu =
1.5 \; 10^{-4} {\rm m}^2/{\rm s}$. The particles trajectories are
resolved in time by a second-order Runge-Kutta scheme, and
interpolated using cubic spline functions. In the dynamic case, a
set of 10,000  particles, uniformly distributed in the cube at
initial time, are followed for a duration of approximatively
7$T_L$.  Here, as in~\cite{ChevPRL}, $T_L$ is defined for each signal as the
time scale above which the velocity statistics  is Gaussian (the
second order cumulant has reached the Gaussian value $\pi^2/8$).
For the static case, 10,000 trajectories have been integrated over
approximatively 3$T_L$. Finally, 32,768 virtual probes have been
used to get the time variation of the Eulerian field at a fixed
location; in this case the records are 10 $T_L$ long. We label
$v_{D,i} (t)$ the Lagrangian velocity of one component (in
cartesian coordinates) of particle number $i$ in the 3D Eulerian
time-evolving flow (dynamic case), $v_{S,i}(t)$ one component of
the velocity advected by the static Eulerian flow (static case)
and $v_{T,i}(t)$ the time evolution of one component of an
Eulerian  velocity probe.

\begin{figure}[t]
\center{\includegraphics[width=7cm]{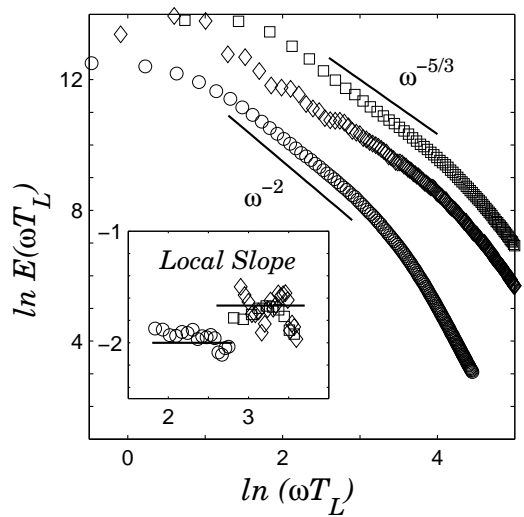}} \caption{Power
spectral density $E(\omega T_L)$ of one component of the velocity
of  tracer particles in the static ($\square$) and  dynamic
($\circ$) cases, and for the virtual Eulerian probes ($\diamond$).
Spectra have been shifted vertically for clarity.  The local
slopes of these spectra are plotted in the inset. }
\label{fig:spec}
\end{figure}

In Fig.~\ref{fig:spec} we show the power spectral densities
$\langle|\hat{v}_S(\omega)|^2\rangle$,
$\langle|\hat{v}_D(\omega)|^2\rangle$  and
$\langle|\hat{v}_T(\omega)|^2\rangle$ vs. $\omega T_L$ in a
logarithmic representation, where $\hat{v}$ means Fourier
transform of $v$, for the static, the dynamic cases (averaged over
the number of tracked particles) and for the virtual fixed probes
(averaged over the number of probes). One observes for all cases,
a scaling behavior on a small range of scales. We show in the
inset the values of the corresponding power law exponent
determined from the local logarithmic slope of the spectra. The
dynamic Lagrangian velocity spectrum has an $\omega^{-2}$ inertial
range spectrum, as expected from Kolmogorov similiraty
arguments~\cite{TenLum}, and in agreement with previous
experimental and numerical observations~\cite{MordantPinton1}. The
spectrum of the time variation of the Eulerian velocity field (as
recorded by the virtual probes) shows a clear $-5/3$ scaling
exponent, characteristic of Eulerian data~\cite{Frisch}. This is
in good agreement with Tennekes sweeping
argument~\cite{Tenn,Vergass}: the characteristic velocity
fluctuations at scale $\ell$ is the standard deviation of the flow
velocity $v_{\rm rms} = \sqrt{\langle|v_S|^2\rangle}$ rather than
the Kolmogorov one $v_\ell = (\epsilon \ell)^{1/3}$.  As a result,
a time-scale increment $\tau$ corresponds to a length $\ell =
v_{\rm rms} \tau$, so that the scaling of the increments is in
fact Eulerian. In addition, this effect produces a larger inertial
range because the dissipative time scale is $T_L R_\lambda^{-3/2}$
for the Eulerian time data, while it is $T_LR_\lambda^{-1}$ for
the dynamic Lagrangian data. Fig.~\ref{fig:spec} then shows that
the data in the static case also follow a $-5/3$ scaling law,
closer to the Eulerian behavior than to the Lagrangian one. This
result has also been found in a similar study using Kinematic
Simulations of turbulence~\cite{Vass}, a situation in which the
Eulerian velocity field is monofractal, {\it i.e.} not
intermittent. Let us mention that since inertial range statistics
are likely to be independent on the Reynolds number, we consider
the behavior of our estimators  in the inertial range (i.e, power
spectra in Fig. 1 and cumulants of magnitude in Fig. 2) as
characteristics of fully developed turbulence --- despite the
moderate Reynolds number value of our DNS ($R_\lambda \simeq
140$).

We now seek to quantify the intermittency features. This is
usually done {\it via} the analysis of the scaling behavior of
velocity structure functions~\cite{Frisch} $S(p,\tau) = \langle
|\delta_\tau v|^p\rangle = \langle |v(t+\tau)-v(t)|^p \rangle \sim
\tau ^{\zeta_p}$, where the average is computed over all
accessible times $t$ and over all recorded time series. Note that
we use the absolute value of velocity increments in the definition
of the structure functions in Eq. (\ref{eq:linkstruccum}) since
the statistics that are studied are symmetric under the
transformation $\delta_\tau v \rightarrow -\delta_\tau v$ :
velocity increment statistics are not skewed. However, as
advocated in~\cite{DelMuz} for Eulerian velocity data
analysis, the magnitude cumulant analysis provides a more reliable
alternative to the structure function method. The relationship
between the moments of $|\delta_\tau v|$ and the cumulants
$C_n(\tau)$ of $\ln |\delta_\tau v|$ reads
\begin{equation}\label{eq:linkstruccum}
\langle |\delta_\tau v|^p\rangle = \exp \left( \sum_{n=1}^\infty
C_n(\tau)\frac{p^n}{n!} \right) \mbox{ .}
\end{equation}
Previous studies \cite{MordantPinton1,MordantPinton2,ChevPRL} have
shown that intermittency is suitably described for small $p$ using
a log-normal statistical framework corresponding to a quadratic
$\zeta_p$ spectrum :
\begin{equation}\label{eq:zetapLN}
\zeta_p = c_1p-c_2\frac{p^2}{2}\mbox{ ,}
\end{equation}
where the parameters $c_1$ and $c_2$ can be extracted from the
time scale behavior of the first two cumulants $C_1(\tau)$ and
$C_2(\tau)$. In  Eulerian context, the analysis of longitudinal
velocity increments in the inertial range has shown that
$C_2^E(\ell) = - c_2^E\ln (\ell/L)$, where $\ell$ is a spatial
scale and $L$ the decorrelation length, with a universal
intermittency coefficient $c_2^E \approx 0.025$ \cite{DelMuz,Cas}.

\begin{figure}[t]
\center{\includegraphics[width=7cm]{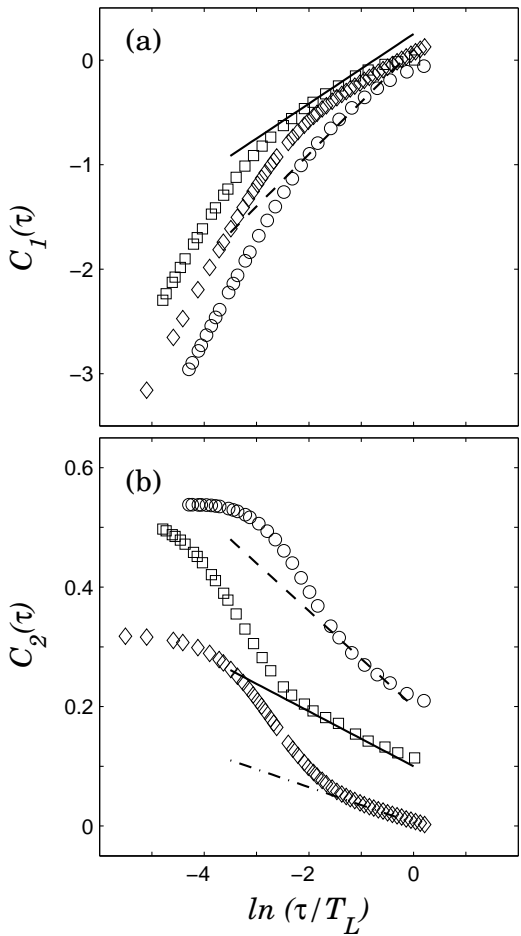}}
\caption{\label{fig:cum} Cumulants of velocity magnitude for the
static ($\square$) and  dynamic ($\circ$) cases, and for the
virtual Eulerian probes ($\diamond$). (a) $C_1^{S,D,T}(\tau)$ vs.
$\ln(\tau/T_L)$; the solid and dashed lines correspond to the
slopes $c_1^{S,T}=1/3$ and $c_1^D=1/2$ respectively; for the sake
of clarity, we have substracted $\langle \ln |\delta_{T_L}v|
\rangle$. (b) $C_2^{S,D,T}(\tau)$ vs. $\ln(\tau/T_L)$; the solid,
dashed-dotted and dashed lines correspond to the slopes
$c_2^S=0.046$, $c_2^{T} = 0.03$ and $c_2^D=0.08$ respectively; we
have subtracted $\mbox{Var}[\ln |\delta_{T_L}v| ] = \pi^2/8$ and
also shifted the upper curves  by 0.1 and 0.2 for clarity.}
\end{figure}

We report in Fig.~\ref{fig:cum} the results of a comparative
analysis of the cumulants $C_1^{S,D,T}(\tau)$ and
$C_2^{S,D,T}(\tau)$. The profile of $C_1^D(\tau)$ as a function of
$\ln(\tau/T_L)$ is significantly curved,  a feature that has also
been observed on experimental data of Lagrangian velocity
structure functions \cite{MordantPinton1}. This departure from
scaling is a signature of (i) the pollution of the inertial range
by dissipative (finite Reynolds number) effects as studied
in~\cite{ChevPRL} and (ii) the non universal and/or anisotropic
behavior of the turbulent flow at scales of the order of the
integral time scale $\sim T_L$. In Fig.~2(a), we have indicated by
a dashed line the scaling behavior $C_1^{D}(\tau) =
c_1^D\ln(\tau/T_L)$, with $c_1^D = \langle h \rangle \simeq 1/2$
corresponding to a $(\omega T_L)^{-2}$ scaling region in the power
spectrum (we recall that $h\simeq1/2$ corresponds to the most
probable velocity scaling exponent in a multifractal analysis of
Lagrangian intermittency~\cite{MordantPinton2,ChevPRL}). In
contrast, the first order cumulant for the static case and for the
Eulerian velocity time variations is better represented by
$C_1^{S,T}(\tau) = c_1^{S,T}\ln(\tau/T_L)$,  with $c_1^{S,T} =
\langle h \rangle \approx 1/3$, as indicated by the continuous
line in Fig. \ref{fig:cum}(a). Again, this value is consistent
with the $(\omega T_L) ^{-5/3}$ power spectrum behavior observed
in Fig.~\ref{fig:spec}. As it is also the case for  Eulerian
fields~\cite{DelMuz,Cas}, the second order cumulants ---
Fig.~\ref{fig:cum}(b) --- has a logarithmic behavior in the
inertial range, $C_2^{S,D,T}(\tau) = - c_2^{S,D,T} \ln(\tau/T_L)$.
In the dynamic case, we get $c_2^D \simeq 0.08$, in good agreement
with previous experimental data~\cite{ChevPRL} and numerical
simulations~\cite{Biferale}.

The static case and the
Eulerian time probes follow the same behavior, with $c_2^{S}
\simeq 0.046$ and $c_2^{T} \simeq 0.03$.  The first value is in
good agreement  with the recent estimate of the intermittency
coefficient $c_2^{E,3D}=0.049$ of a numerical 3D Eulerian velocity
field~\cite{Kestener}. Note that $c_2^{E,3D}$ is significantly
larger than the value $c_2^{E} \simeq 0.025$ computed for
longitudinal velocity increments, to which one should  compare the
estimate of the intermittency coefficient $c_2^{T}$ of the time
variations of the Eulerian  longitudinal velocity component. As a
technical but important point, we stress that the values of $c_2$
reported here are not computed from a linear regression fit in the
$C_2(\tau)$ curves (due to the narrowness of the inertial range)
but as a result of the mutifractal description of the entire range
of scales, dissipative domain included, as detailed in~\cite{ChevPRL}.

\begin{figure}[t]
\centerline{\includegraphics[width=7cm]{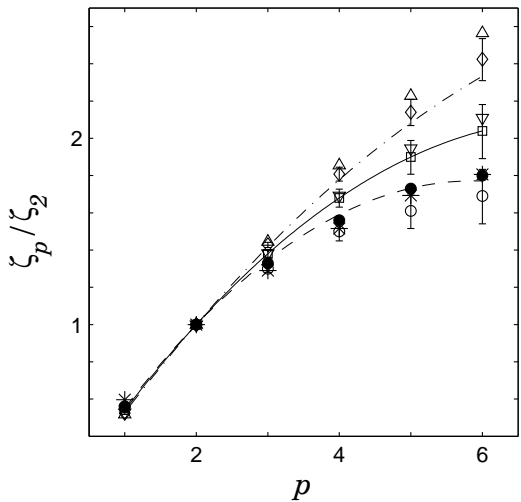}}
\caption{\label{fig:zetap} Structure function exponent
$\zeta_p/\zeta_2$ vs. $p$ as computed for the static case
($\square$), the dynamic ($\circ$) case and the Eulerian time
probes ($\diamond$). Are also shown for comparison the exponents
obtained for Lagrangian velocity experiments ($\bullet$)
\cite{MordantPinton1}, 3D Eulerian velocity fluctuations obtained
by DNS ($\nabla$) \cite{Kestener}, experimental 1D Eulerian
longitudinal velocity increments ($\triangle$) \cite{Cas} and
passive scalar increments ($\ast$) \cite{Passive}. The solid,
dashed-dotted and dashed lines correspond to the quadratic $\zeta_p$
spectra (Eq. (2)) with the parameter values: $(c_1^{S}=1/3; \;
c_2^{S}=0.046)$, $(c_1^{T}=0.3; \; c_2^{T}=0.03)$, and
$(c_1^{D}=1/2; \; c_2^{D}=0.085)$. }
\end{figure}

We thus conclude that the static and Eulerian intermittencies in
time have identical statistics to respectively 3D and 1D Eulerian
fields, while the true dynamic (Lagrangian) case is clearly
different and more intermittent. This is confirmed in
Fig.~\ref{fig:zetap} where the more familiar structure function
exponents $\zeta_p$ are compared. Note that in order to be able to
compare Eulerian and Lagrangian data, all exponents are computed
using the second order structure function as a reference (extended
self-similarity  (ESS) ansatz \cite{ESS}) in order (i) to overcome
the observed bending of the structure functions when plotted
versus the time scale in a logarithmic representation (as already
noticed in Fig. \ref{fig:cum}(a) for the first cumulant
$C_1(\tau)$), and (ii) to give a clear picture of intermittency
effects responsible for departure from a linear behavior (i.e.
$\zeta_p^D \ne p/2$ and $\zeta_p^T, \zeta_p^S \ne p/3$). Thus, in
this representation, each $\zeta_q$ spectrum must be compared to
the monofractal Kolmogorov's prediction $\zeta_p=p/2$.
As the size of the statistical ensemble available is limited, we
restrict ourselves to moments of order $p$ up to 6. The values
obtained are in agreement with a parabolic spectrum (Eq.
(\ref{eq:zetapLN})) when fixing the parameters $c_1$ and $c_2$ to
the values previously estimated from the magnitude cumulants and
auto-correlation functions. In Fig. \ref{fig:zetap} are also shown
for comparison the $\zeta_p$ spectra obtained for experimental
Lagrangian velocity measurements, experimental Eulerian
longitudinal increments, and full 3D numerical Eulerian velocity
multifractal analysis. Once again, the static and Eulerian time
behavior are identical respectively to that of the 3D numerical
Eulerian velocity~\cite{Kestener} and to the traditional 1D
longitudinal velocity increments~\cite{Cas}. But they are both
less intermittent than observed for the dynamic case, {\it i.e.}
for the Lagrangian velocity field. A detailed account of the
relationship between Eulerian and Lagrangian intermittencies, in
the framework originally proposed by M. Borgas~\cite{Borgas} has
been previously discussed in~\cite{ChevPRL}. Finally, we note that
another quite noteworthy feature in Fig. \ref{fig:zetap} is that
the exponents for the passive scalar increments are identical to
that of the Lagrangian velocity statistics (within error bars).

To summarize our findings, we have observed that the statistics of
particles advected in a frozen Eulerian field is, to some extent,
similar to that of the time variations of the Eulerian velocity at
fixed points in space. This is an ergodicity property of
homogeneous, isotropic turbulence. The similarity of the static
case and the full 3D Eulerian field could prove useful because
local variations in time are way easier to measure than full
spatial 3D flows --- although it is absolutely necessary that the
mean flow be truly absent, otherwise local time variations relate
to spatial profiles, using Taylor's
hypothesis~\cite{TenLum,Tayloc}. Finally, the intermittency
measured in the dynamics case is different, showing that true
Lagrangian particles are sensitive to the global  time evolution
of the flow. One eventually expects that the large scale dynamics
is even more crucial in understanding mixing effects in real
non-homogeneous flows.

This work is supported by the French Minist$\grave{\mbox{e}}$re de
la Recherche (ACI) and the Centre National de la Recherche
Scientifique under GDR Turbulence. Numerical simulations are
performed at CINES (France) using an IBM SP computer.


\end{document}